\documentclass{PoS}

\let\ifpdf\undefined
\usepackage{url}
     
\title{Recent results of the ATLAS Upgrade Planar Pixel Sensors R\&D Project}

\ShortTitle{Recent results of the ATLAS Upgrade Planar Pixel Sensors
  R\&D Project}

\author{\speaker{Giovanni Marchiori}\\
        for the ATLAS Upgrade Planar Pixel Sensors Collaboration\\
        LPNHE Paris\\
        E-mail: \email{giovanni.marchiori@lpnhe.in2p3.fr}}

\abstract{
  The ATLAS detector has to undergo significant updates at the end of 
  the current decade, in order to withstand the increased occupancy and 
  radiation damage that will be produced by the high-luminosity upgrade of the
  Large Hadron Collider.
  In this presentation we give an overview of the recent
  accomplishments of the R\&D activity on the planar
  pixel sensors for the ATLAS Inner Detector upgrade.}

\FullConference{10th International Conference on Large Scale
  Applications and Radiation Hardness of Semiconductor Detectors,\\ 
		July 6-8, 2011\\
		Firenze Italy}

\begin{document}

\section{Introduction}

%LHC
The Large Hadron Collider (LHC) at CERN is a proton-proton collider
designed to reach an unprecedented high center-of-mass energy
($\sqrt{s} = 14$~TeV) and luminosity ($10^{34}$~cm$^{-2}$s$^{-1}$).
It operates smoothly since spring 2010 at a center-of-mass energy of 7
TeV and a luminosity which has constantly increased up to around half
the design value, and is expected to attain the nominal performance in
the years following the shutdown planned for 2013 and 2014.
%ATLAS
ATLAS (A Toroidal LHC ApparatuS) is a general purpose detector built
around one LHC interaction point in order to reconstruct the
products of the highly energetic $pp$ collisions~\cite{bib:ATLAS_detector}.
Its primary goals are to probe the Standard Model predictions at high
energy, to discover the Higgs boson and to search for beyond-Standard
Model evidences, like supersymmetric particles or other exotic processes.
%ATLAS ID and pixel
The Inner Detector (ID) is the ATLAS subdetector closest 
to the interaction point.
It is made of three nested subsystems immersed in a 2 T
axial magnetic field: a silicon pixel
detector~\cite{bib:pixel_detector} at small radial distance $r$ from
the beam axis ($50.5{<}r{<}150$ mm), double layers of single-sided
silicon microstrip detectors ($299{<}r{<}560$ mm), and a straw tracker
with transition radiation detection capabilities ($563{<}r{<}1066$
mm). The ID allows an accurate reconstruction of
tracks from the primary $pp$ collision region and from 
secondary vertices due to photon conversions or heavy flavor decays.

%LHC HL update
To extend the physics reach of the LHC~\cite{bib:LHC_upgrade},
a two-step upgrade of the accelerator
is planned which will increase the peak luminosity by a factor 5 to 10
towards the end of this decade.
The first step ({\em phase I}, around 2017) will require relatively
small upgrades (linac, collimation) and will bring the LHC luminosity
to $2-3\times 10^{34}$ cm$^{-2}$s$^{-1}$. 
The second step ({\em phase II}, around 2021) will require a major
effort (new final focusing design) and will raise the luminosity
further up to $5\times 10^{34}$~cm$^{-2}$s$^{-1}$ (including
luminosity leveling).

\begin{figure}[!htbp]
\centering
\includegraphics[width=0.45\textwidth]{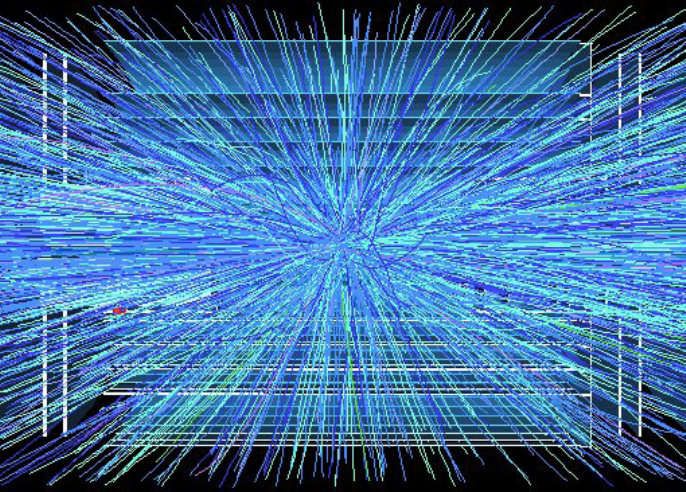}
\hspace{0.03\textwidth}
\includegraphics[width=0.5\textwidth]{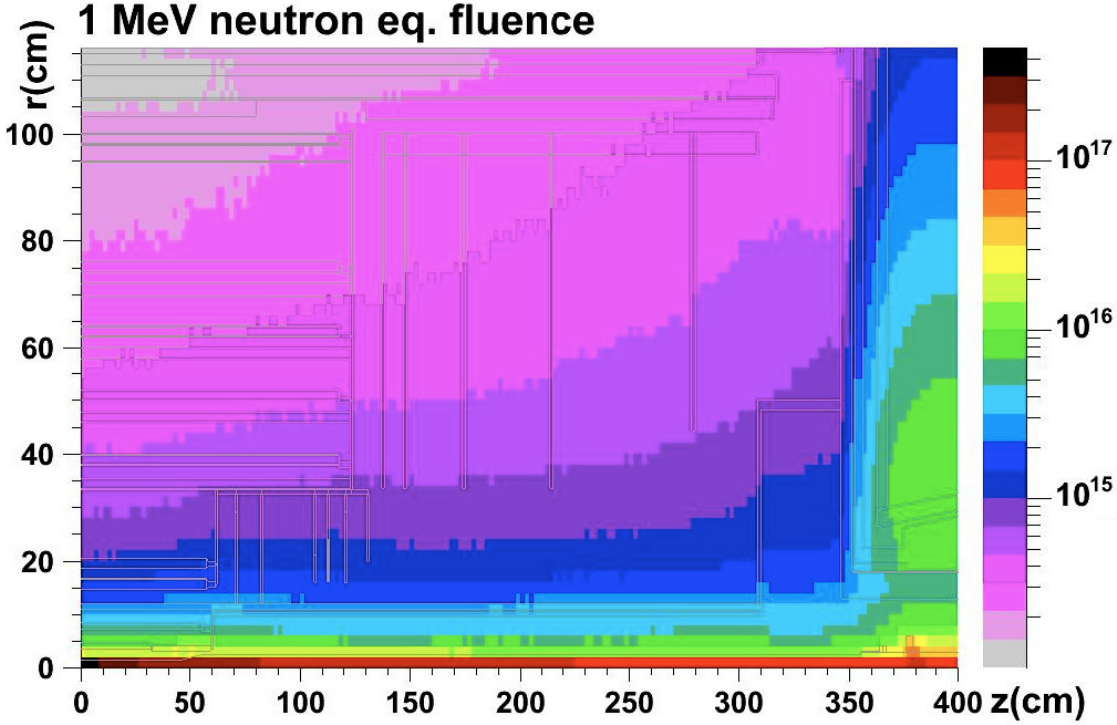}
\caption{Left: pile-up events at a luminosity of $1\times 10^{35}$
  cm$^{-2}$s$^{-1}$. Right: fluences in n$_{\rm eq}/{\rm cm}^2$
  expected for a phase II integrated luminosity of 3000 fb$^{-1}$. A
  safety factor of 2 is applied.}
\label{fig:slhc_radiation_effects}
\end{figure}

The ATLAS pixel detector is composed of 250 $\mu$m thick n-in-n
DOFZ-Si pixel sensors with cell sizes of $50\times 400$~$\mu$m$^2$,
readout by the DC-coupled (bump-bonded) FEI3 chips~\cite{bib:FEI3}.
It has been designed to withstand a fluence of $10^{15}$ ${\rm
  n}_{\rm eq}/{\rm cm}^2$, which will be exceeded in the inner layer before 
the end of phase I. For this reason,
a fourth pixel layer (IBL~\cite{bib:IBL}) at $r=3.7$ cm will be inserted
during the 2013-2014 shutdown and is designed to withstand a total fluence of
$2\times 10^{15}~{\rm n}_{\rm eq}/{\rm cm}^2$ ($5\times 10^{15}~{\rm n}_{\rm eq}/{\rm cm}^2$ including safety factors).
The whole pixel detector will be replaced before the start of phase II,
in order to cope with the increased
occupancy (between 100 and 200 pile-up collisions, 
an order of magnitude higher than at the nominal LHC
luminosity) and radiation damage, with the innermost layer being
exposed to a total fluence of $2\times 10^{16}~{\rm n}_{\rm eq}/{\rm cm}^2$
(Fig.~\ref{fig:slhc_radiation_effects}).

%PPS Collaboration
\section{The ATLAS Upgrade PPS Collaboration}
The LHC upgrade poses several challenges to the ATLAS pixel detector:
\begin{itemize}
\item radiation damage: the much higher fluences imply increased
  leakage currents, a higher bias voltage needed to (at least
  partially) deplete the devices, and larger charge trapping
  (lower charge collection efficiency). S/N will decrease.
\item data rate: the higher interaction rate requires smaller pixels
  to keep occupancy under control. Pixels of size $50\times
  250$~$\mu$m$^{2}$ are being produced for the IBL, even smaller
  pitches may be used for the inner layers for the phase II upgrade. 
  For the same reason, the replacement of the innermost short strip
  layer(s) by pixel layer(s) might be desirable, however, a
  significant cost reduction of pixel sensors is necessary to 
  compete with strip sensor costs.
\item tight geometrical constraints: with the addition of a
  fourth pixel layer closer to the beampipe, the limited space between
  the first layers will not allow the "shingling" of the detector
  modules along the beam direction. In addition, flat 
  staves ease cooling and production.
  To minimize the geometric efficiency losses the edges of the modules
  have to be active or at least slimmer ($<450~\mu$m) than the current
  ones. The inactive edge of the present modules is 1.1~mm, hosting 16
  guard rings (total width $\approx 600~\mu$m) to manage the potential
  drop between the cutting edge of the sample and the pixel
  electrodes, and an additional safety margin of 500~$\mu$m to
  insulate the guard ring area from damages of the lateral surface due
  to the dicing. 
\end{itemize}
To investigate the suitability of pixel sensors using the proven
planar technology, already exploited successfully in ATLAS and other
high-energy-physics (HEP) experiments at colliders, for the 
upgraded tracker, the ATLAS Planar Pixel Sensor R\&D Project was
established~\cite {bib:PPS_proposal}.
It is formed by more than 80 scientists from 17 European, US and Japanese 
institutes. 
Main areas of research are the performance of planar pixel sensors at
phase I and II fluences, the exploration of possibilities for cost
reduction to enable the instrumentation of large areas, the
achievement of slim or active edges to provide low geometric
inefficiencies without the need for shingling of modules and the
investigation of the operation of highly irradiated sensors at low
thresholds to increase the efficiency. 
Several test prototypes have been produced at various foundries (CiS,
Micron, Hamamatsu, MPI-HLL) in collaboration with RD50~\cite{bib:RD50} and
characterized by probe station (IV/CV curves), with radioactive
sources, and in testbeams with pions at CERN and electrons at DESY (hit
efficiency, charge collection), before and after irradiations. 
Both neutron irradiation (with the Triga reactor in Ljubljana) and
proton irradiation (using a 23 MeV/c beam in Karlsruhe and the
24 GeV/c beam of the CERN PS) have been performed.
The optimization of some of the sensor parameters, for instance the
geometrical guard ring (GR) configuration (number of GRs, widths,
distances), has been supported by the use of TCAD electrical simulation
tools~\cite{bib:silvaco, bib:hiroshima}.
In the following sections we summarize the
most recent results of these ongoing R\&D studies.

\section{Radiation hardness of $n$-type and $p$-type pixel sensors}
Only electron-collecting sensors designs (n-in-p and n-in-n) have been 
considered in the PPS R\&D, for two radiation damage related reasons: 
the hole trapping time is unacceptably short at large fluences, 
and p-in-n (after type inversion) and p-in-p sensors cannot be operated
efficiently at partial depletion because of the undepleted volume near the 
electrodes.

\begin{figure}[!htbp]
\centering
%\includegraphics[width=0.45\textwidth]{cce_ninn_5e15}
%\hspace{0.03\textwidth}
%\includegraphics[width=0.45\textwidth]{cce_ninn_2e16}
\includegraphics[width=0.9\textwidth]{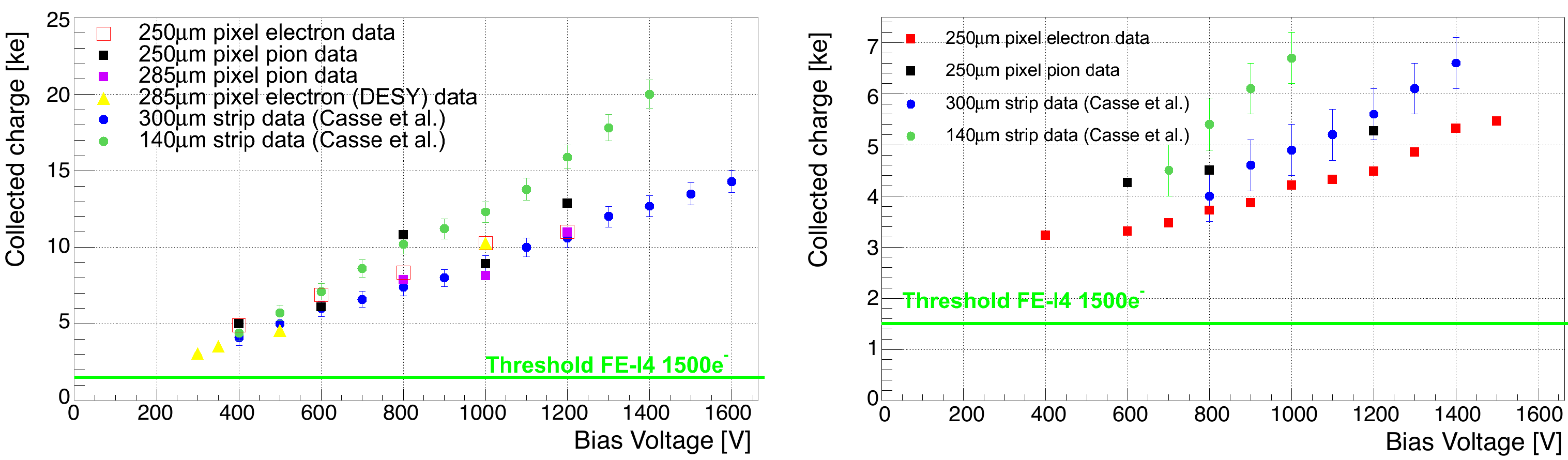}
\caption{Charge collection vs bias voltage for n-in-n pixel and strip
  sensors at fluences of $5\times 10^{15}$ n$_{\rm eq}/{\rm cm}^2$
  (left) and $2\times 10^{16}$ n$_{\rm eq}/{\rm cm}^2$ (right). Strip
  sensor data are from~\cite{bib:casse}.} 
\label{fig:cce_ninn}
\end{figure}

\begin{figure}[!htbp]
\centering
\includegraphics[width=0.83\textwidth]{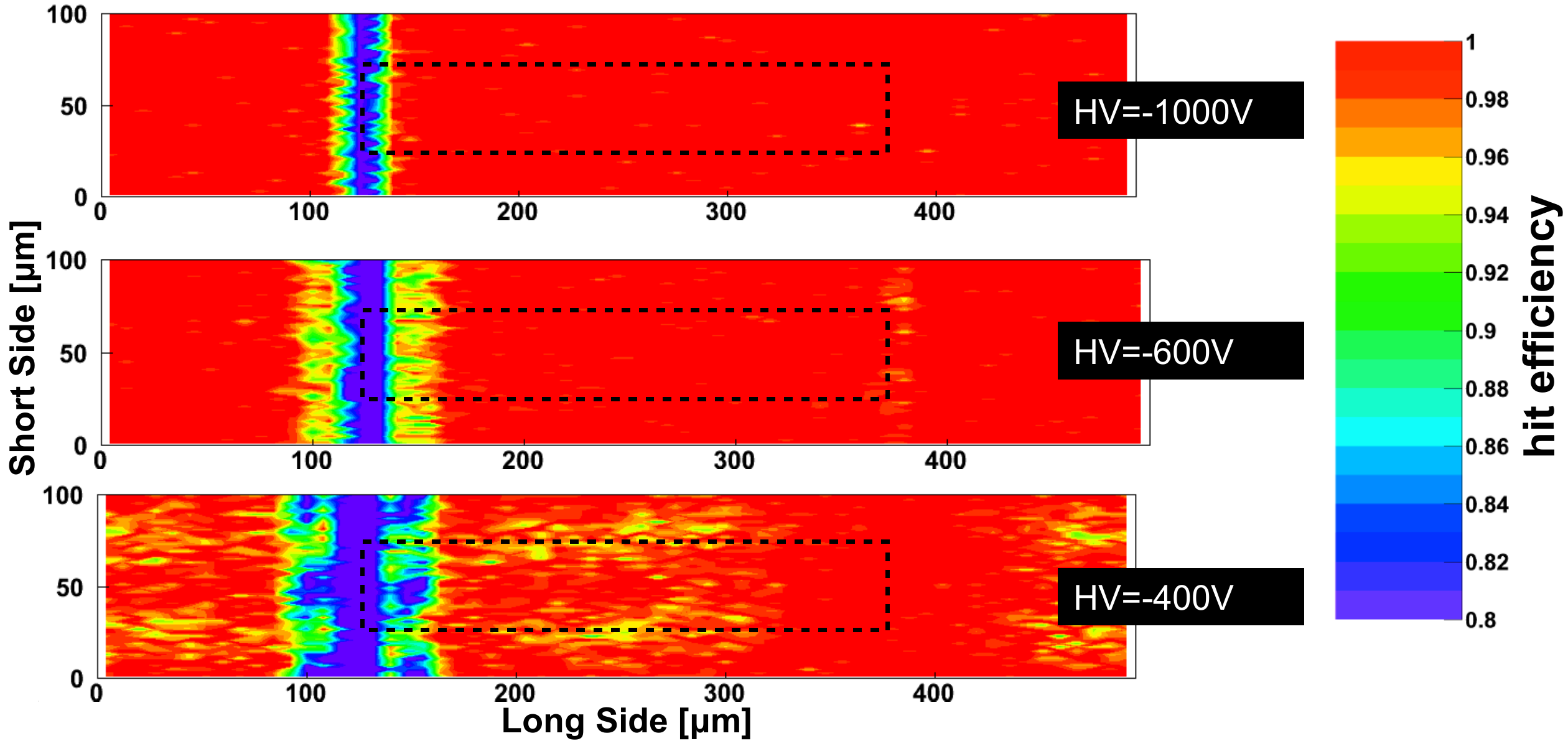}
\caption{Hit efficiency as a function of the hit position in 250 $\mu$m thick,
n-in-n sensors with $50\times 250~\mu{\rm m}^2$ pixel cells, irradiated
with neutrons at an equivalent fluence of $4\times 10^{15}$ 
n$_{\rm eq}/{\rm cm}^2$. 
The dashed rectangle ($125<x<375~\mu$m, $25<y<75~\mu$m) corresponds to one full 
pixel cell, while only half pixels are shown for neighbouring pixels. 
The low-efficiency vertical region near $x=125~\mu$m corresponds to the 
bias line between two pixel columns.}
\label{fig:hiteff_fei4_cerntb}
\end{figure}

The radiation hardness of n-in-n prototypes has been tested after
neutron irradiations up to fluences of $2\times 10^{16}$ n$_{\rm
  eq}/{\rm cm}^2$. Operating temperatures of $-30$~C have been chosen
to reduce the leakage current and therefore shot noise. Testbeam data 
on devices connected to readout chips tuned for thresholds of 1500~e show that 
about half (10~ke) of the charge deposited by a MIP in 250 $\mu$m of
Si can be collected with a bias voltage of 1~kV, which is the maximum
allowed for the IBL at the end of the LHC phase I, after $5\times 10^{15}$ n$_{\rm eq}/{\rm cm}^2$
(Fig.~\ref{fig:cce_ninn}, left).
This translates into a signal-over-treshold ratio $S/T>6$ and a
signal-to-noise ratio $S/N>45$. Similar studies show that some charge
(around 5.5~ke) can be collected even at $2\times 10^{16}$ n$_{\rm
  eq}/{\rm cm}^2$ by increasing the bias voltage to 1.5~kV
(Fig.~\ref{fig:cce_ninn}, right). 
Preliminary measurements of the hit efficiency in 250 $\mu$m thick,
$50\times 250~\mu{\rm m}^2$ n-in-n irradiated sensors performed in the
June 2011 testbeam at the CERN SPS~\cite{bib:ibl_june11_tb}
(Fig.~\ref{fig:hiteff_fei4_cerntb}) show that already at $V_{\rm
  bias}=600~V$ an efficiency greater than 97\% can be achieved after
neutron irradiation to $4\times 10^{15}$ n$_{\rm eq}/{\rm cm}^2$.  

\begin{figure}[!htbp]
\centering
\includegraphics[height=4.5cm]{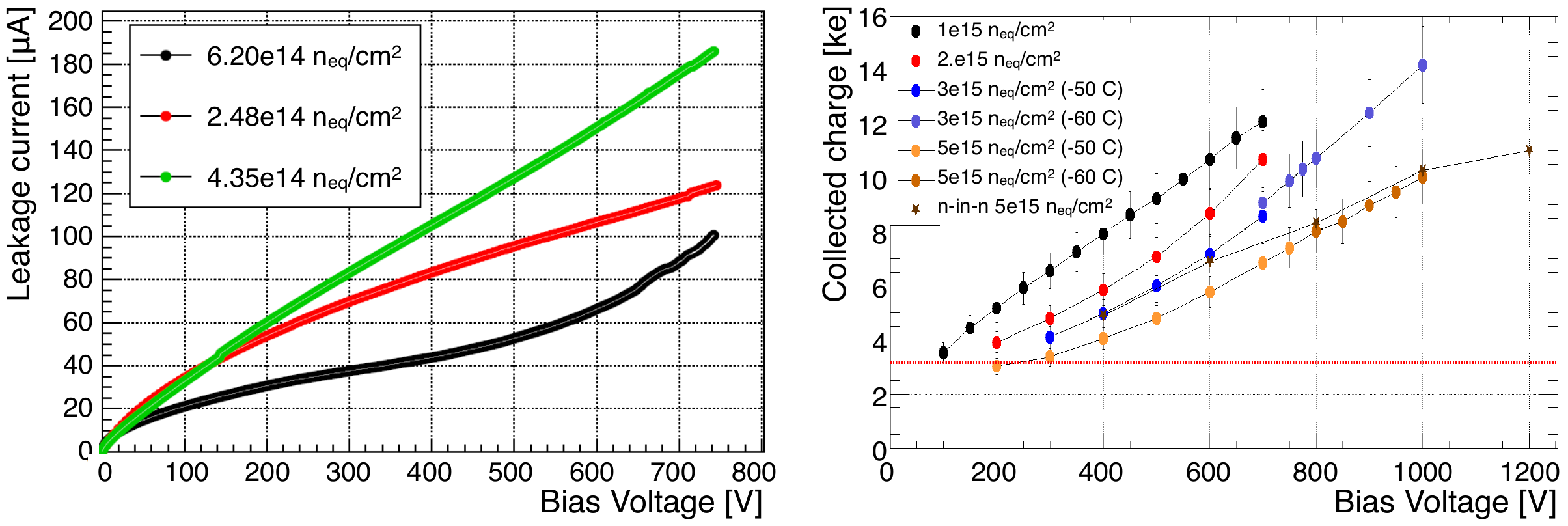}
\caption{IV curves (left) and collected charge vs V$_{\rm bias}$
  (right) at various fluences for n-in-p pixel sensors irradiated with
  neutrons. The horizontal
  red line corresponds to the threshold set in the front-end chip 
  discriminator.}
\label{fig:ninp_5e15}
\end{figure}

In recent years the interest for p-bulk devices as possible sensors for HEP experiments has 
rapidly increased as high-resistivity p-type wafers have become available and since they do not 
undergo type inversion after heavy irradiation.
Moreover, since guard rings are located on the same side as the pixel
electrodes, the layout of these sensors requires a single sided
production process with less steps than for n-in-n sensors, and
possible cost reduction. 
A potential issue is that since the GRs are on the same side as the
pixels, which are connected to the front end electronic chip which is
at ground, the full bias voltage is right at the edge of the sensor
and only a few $\mu$m away from the chip. The high bias voltages
needed to deplete the sensors may thus generate sparks that would
destroy the chip. 
With an additional passivation layer ($\approx$ 3  
$\mu$m) of a highly insulating material like benzocyclobutene or
parylene on top of the sensors this effect has not been observed for
bias voltages up to 1 kV.
Recently 285~$\mu$m thick n-in-p sensors were produced at CiS and
bump-bonded to FEI3 chips at IZM-Berlin.
Prior to irradiation they exhibited breakdown voltage much higher than
the full depletion voltage  (60 V), leakage currents around 500 nA/cm$^2$ 
(unaffected by bump-bonding), narrow threshold tuning with 
a mean of 3200 e and dispersion 22 e and 100\% charge collection efficiency 
at full depletion. S/N ($\approx$ 19) and hit efficiency (>99\%,
measured in testbeams) were similar  to that of n-in-n pixels (and of
n-in-p sensors produced by HPK and Micron). 
These devices have been subsequently tested after irradiation with
either protons or neutrons.
After a fluence of about $5\times 10^{15}$ n$_{\rm
  eq}/{\rm cm}^2$, the breakdown voltage is greater than 750 V and the
leakage current lower than 200 $\mu$A at $-10$ C
(Fig.~\ref{fig:ninp_5e15}, left). A significant fraction (>30\%) of
the charge generated by a MIP is collected at a high bias voltage (>600
V) and yields a detectable signal above threshold
(Fig.~\ref{fig:ninp_5e15}, right).
The charge collection efficiency is similar to that of n-in-n devices 
exposed to similar hadron fluences. 
%% An even larger charge may be collected at higher V$_{\rm bias}$
%% (>1500 V) due to charge amplification as observed in n-in-p sensors
%% previously produced at Micron. 

\section{Sensor thinning}
\begin{figure}[!htbp]
\centering
\includegraphics[width=0.7\textwidth]{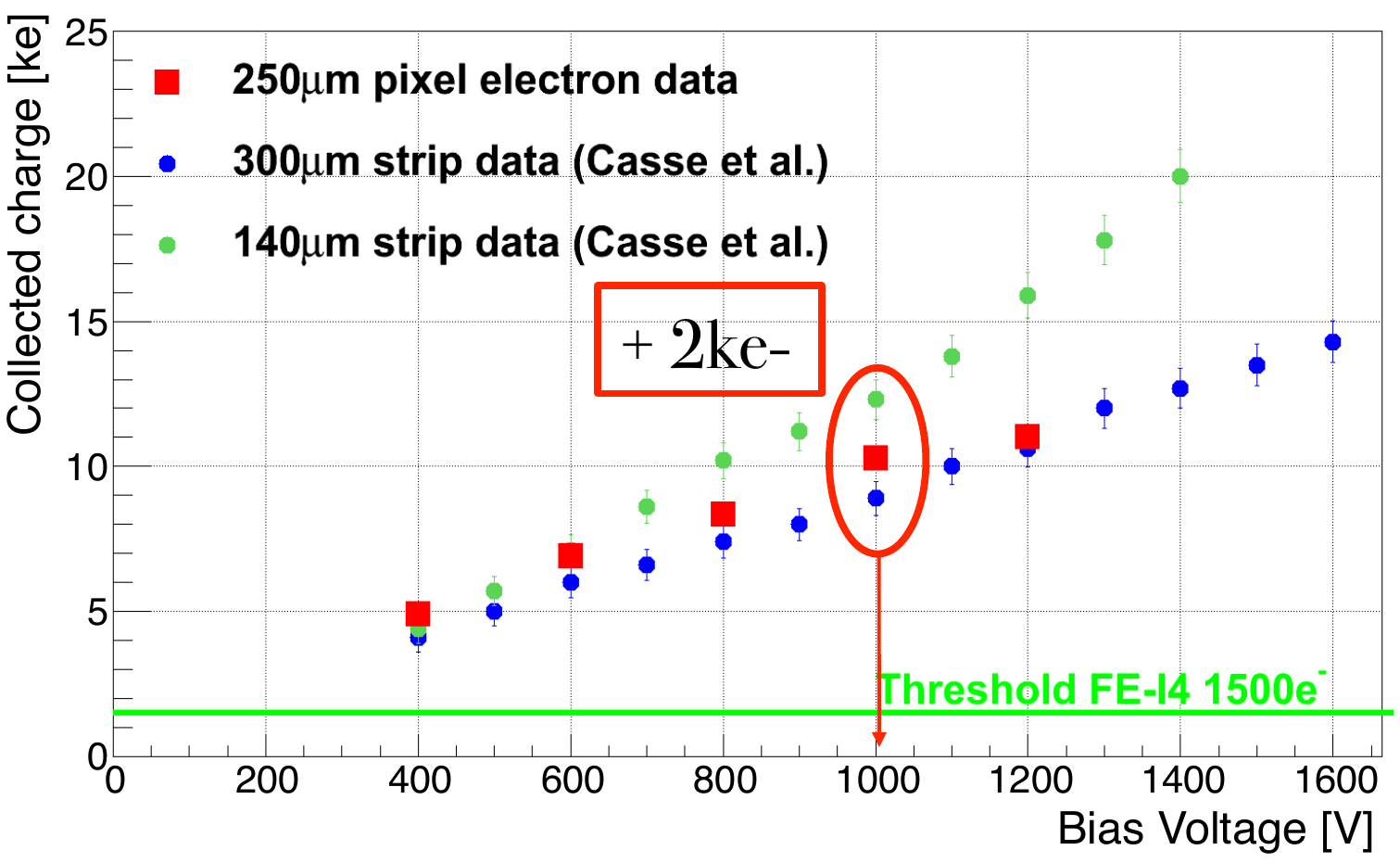}
\caption{Collected charge vs V$_{\rm bias}$ for n-in-n sensors of different 
  thicknesses exposed to a neutron fluence of 
  $5\times 10^{15}$ n$_{\rm eq}/{\rm cm}^2$.
  The horizontal red line corresponds to the threshold
  set in the front-end chip discriminator.}
\label{fig:ninn_thinning_cce}
\end{figure}

Recent strip-sensor data show that thinner sensors yield more charge
than thicker ones after fluences of $5\times 10^{15}$ n$_{\rm eq}/{\rm
  cm}^2$ for bias voltages of 1 kV or higher, because of the higher
electric field and possibly also due to charge multiplication inside the  
bulk~\cite{bib:casse}.
Thinner sensors for HEP experiments have the additional advantage of
the lower material budget, which improves momentum resolution and
reduces energy loss in front of downstream detectors. 
Recently CiS has started to produce for the PPS R\&D n-in-n sensors
with thicknesses between 150 and 250 $\mu$m, in steps of 25
$\mu$m. Part of the 250 $\mu$m sensors have been delivered, with good
wafer yield (90\%) and sensor-on-wafer yield close to 100\%. 
At V$_{\rm bias}=1$ kV they collect more charge than 300 $\mu$m thick strips 
(Fig.~\ref{fig:ninn_thinning_cce}).
Similar studies will be performed on the thinner sensors as soon as
they become available.
Also n-in-p sensors with thicknesses of 150 and 200 microns are being
produced by the same vendor and will be tested in the near future.

\section{Reducing the inactive area with slim-edge or active-edge devices}
To avoid that the depletion region spreading from the active area reaches the damaged lateral surface, an undepleted region is usually present at the sensor's edge.
This inactive area reduces charge collection near the edge. 
To increase the active area we are exploring two techniques.

\begin{figure}[!htbp]
\centering
\includegraphics[width=0.45\textwidth]{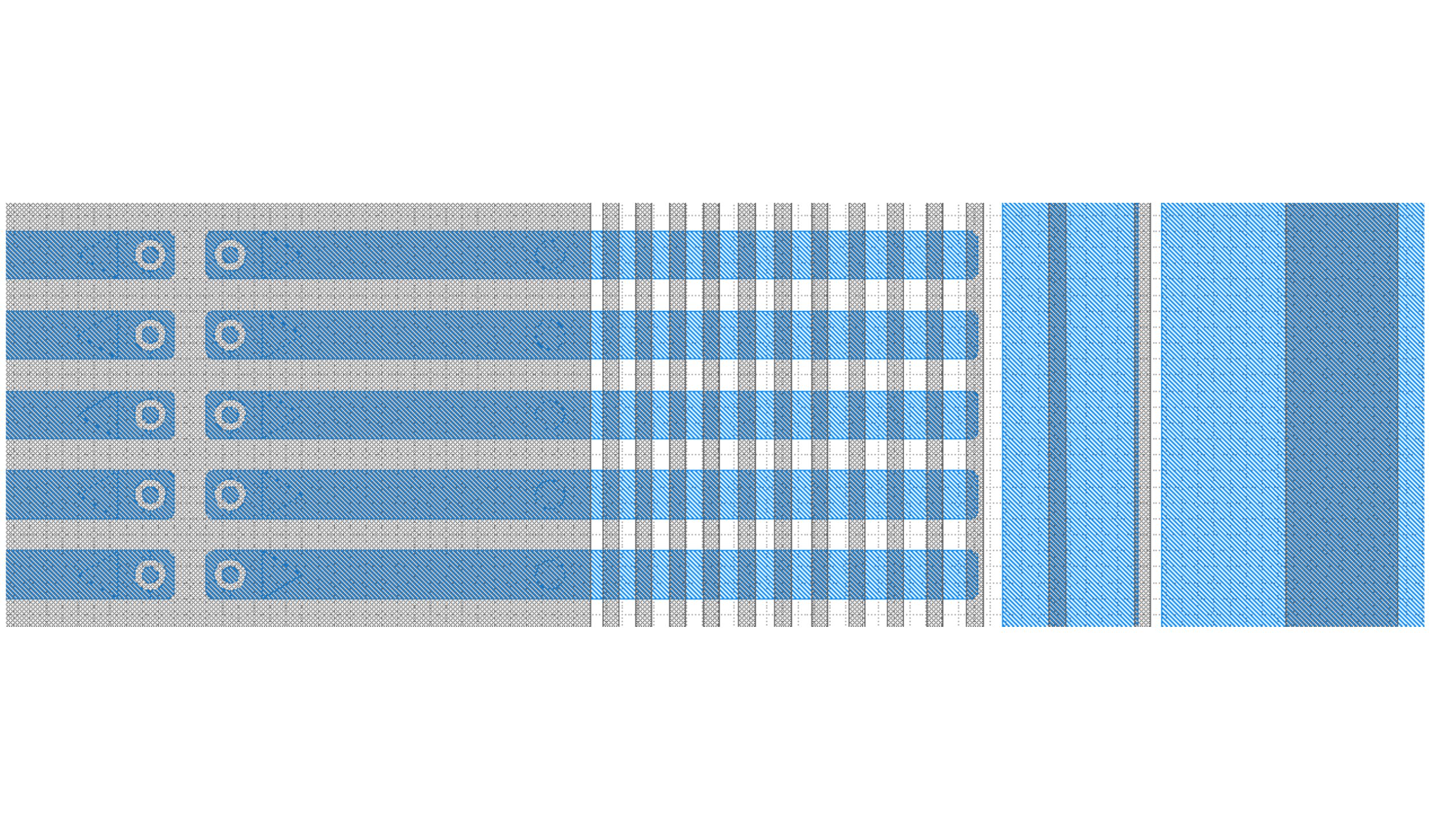}
\hspace{0.03\textwidth}
\includegraphics[width=0.45\textwidth]{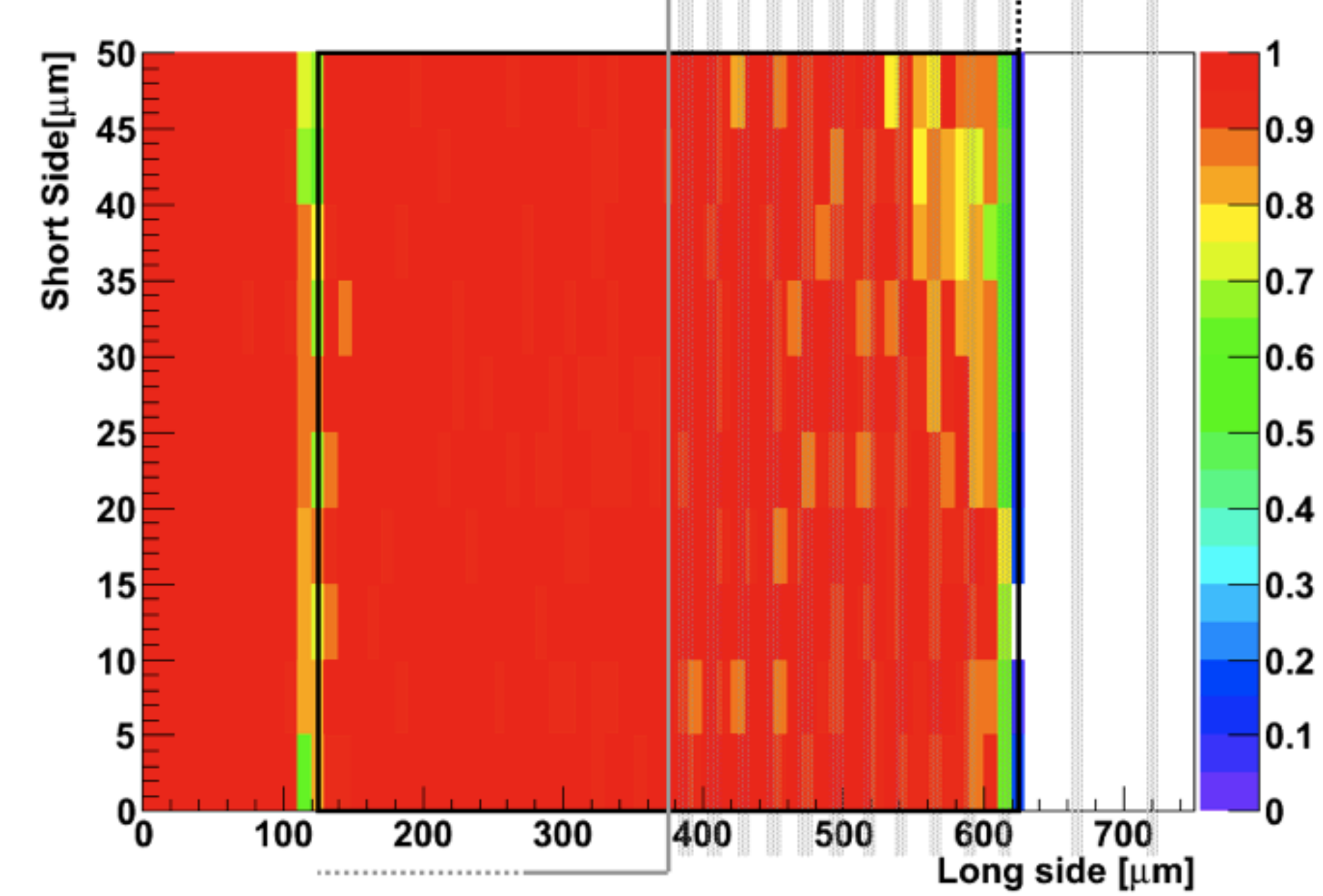}
\caption{Slim-edge sensors. Left: slim-edge design. The pixels
  (horizontal blue rectangles) close to the edge are longer
  (500~$\mu$m) and partially overlap the guard ring region (vertical
  narrow grey rectangles) on the opposite side. Right: hit efficiency 
  map of slim-edge sensors irradiated with neutrons at $4\times 10^{15}$
  n$_{\rm eq}/{\rm cm}^2$ and biased at 1000 V. Data was taken with an
  incidence angle of 15$^\circ$.
  The sensor extends in the region $125<x<625~\mu$m. Bias dots are
  close to $x=125~\mu$m. The guard ring region is located at $x>380~\mu$m.}
\label{fig:slim_edges}
\end{figure}

Edges can be made "slim" by reducing both the width of the guard ring
area, optimizing the number and spacing of the GRs, and the safety margin, exploiting dicing 
techniques that create less surface damage.
Based on simulations and on several dicing trials showing a good
yield, a reduced number of 13 guard rings with a safety margin of 100
$\mu$m was proposed for the IBL sensors. 
A further gain for n-in-n pixels can be obtained by extending the
pixel implants, which are located on the side opposite to the GRs, to
partially overlap the guard rings (Fig.~\ref{fig:slim_edges}, left). 
In the overlap region the bulk is (at least partially) depleted and
sufficient charge is collected, with a reduction of the inactive area 
to less than 1.5\%.
This has been investigated with n-in-n slim-edge sensors in a recent
testbeam at CERN.
Preliminary results show a high efficiency (close to 100\%) up to
about 250~$\mu$m from the cutline for non-irradiated devices as well
as for devices irradiated with neutrons up to $4\times 10^{15}$
n$_{\rm eq}/{\rm cm}^2$ and biased at 1 kV (Fig.~\ref{fig:slim_edges},
right)~\cite{bib:CERN_tb}.
Given the good performance and yields achieved, the slim-edge design
has been chosen as the IBL candidate technology for the planar pixels
and is currently being produced.

\begin{figure}[!htbp]
\centering
\hspace{0.05\textwidth}
\includegraphics[width=0.3\textwidth]{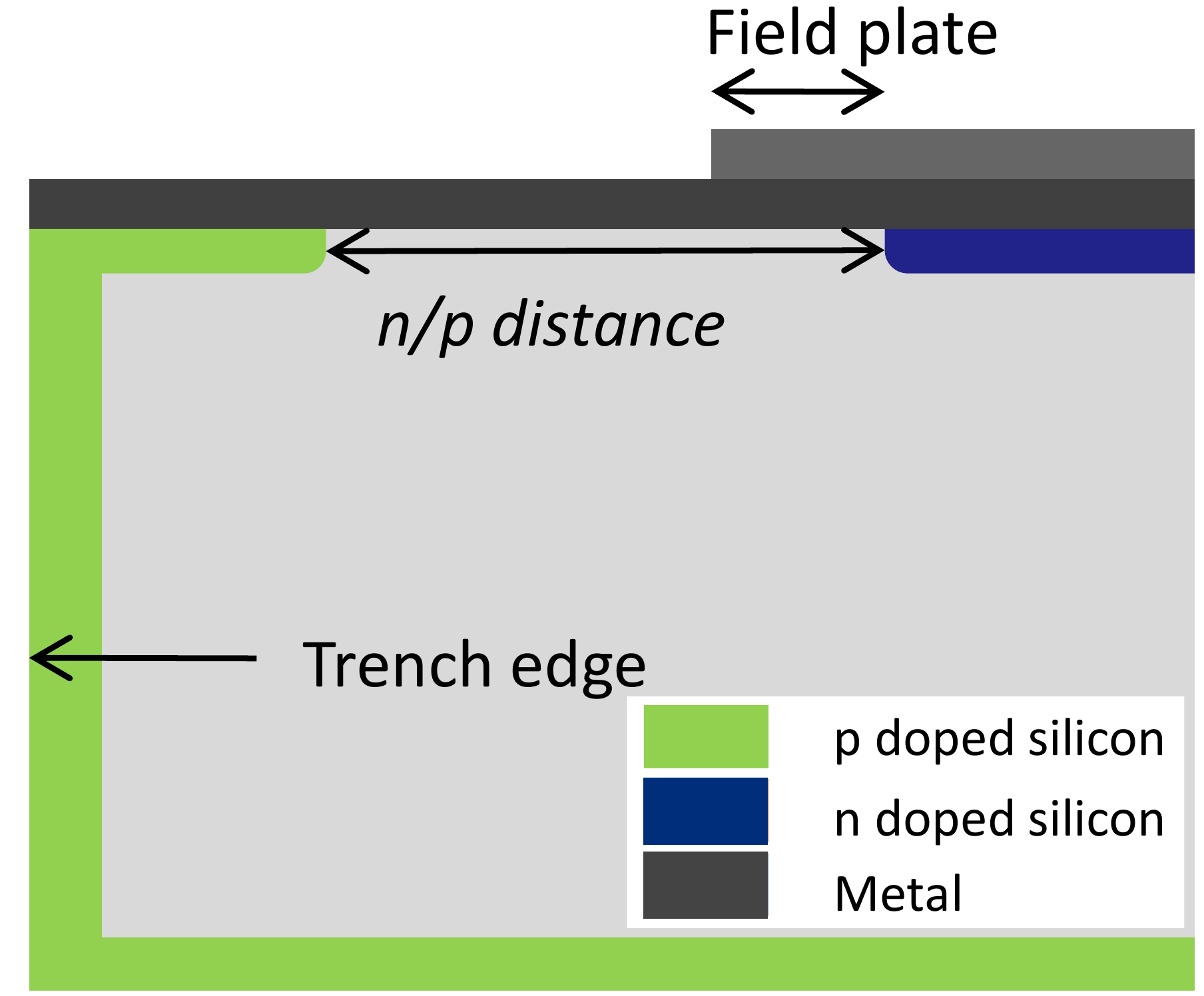}
\hspace{0.15\textwidth}
\includegraphics[width=0.45\textwidth]{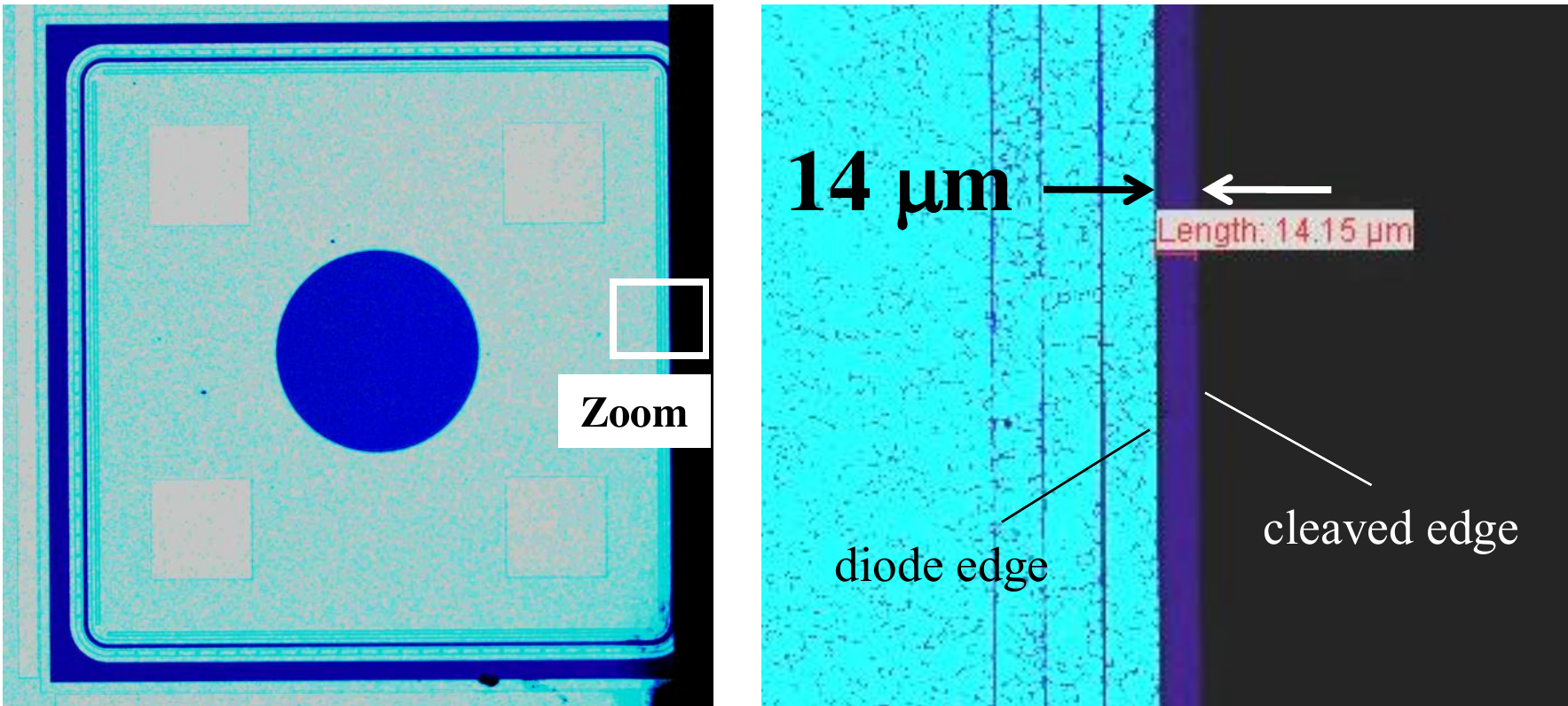}
\caption{Edgeless sensor technologies. Left: active edge design. A p-doped implant (green) is deposited on the lateral surface of the module (in addition to the p HV implant on the back side). Right: a diode after laser dicing of one lateral surface. All guard rings have been removed.}
\label{fig:active_edges}
\end{figure}

A second technique for the reduction of the inactive area is the
active-edge concept. Within the PPS R\&D two technologies are being
investigated. The former, proposed in~\cite{bib:active_edge} and
available at several foundries collaborating with the PPS group (CNM,
FBK, VTT), uses Deep Reactive Ion Etching (DRIE) to obtain deep
trenches around the detectors, thus eliminating the need of the
cutting procedure.
These trenches are heavily doped so as to behave like ohmic contacts
(Fig.~\ref{fig:active_edges}, left).
The fabrication process requires a support wafer and presents
technological challenges in the excavation of deep (200-230~$\mu$m)
trenches with aspect-ratio around 1:20 and in filling the trenches
with polysilicon with good planar surface, as needed for proper
deposition of the photoresist for the later production stages. 
Preliminary measurements on test diodes produced at FBK with this
technique show reasonably low leakage currents and breakdown voltages
higher than the full depletion value.~\cite{bib:active_edge_FBK}.
A production of planar pixel sensors with this technique is currently
being planned. 
The second edgeless technology is based on laser scribing and cleaving, 
which result in significantly less surface
damage~\cite{bib:active_edge_laser} thus eliminating the need for a
safety margin, and on passivation of the lateral surface with either
silicon oxide or nitride (for n-type bulk sensors) or atomic layer
deposition (ALD) of aluminum oxide (for p-type bulk). 
This creates an interface charge (positive for silicon oxide and
negative for aluminum oxide) on the lateral surface that is able, if
its density is sufficiently high (around $10^{12}$ cm$^{-2}$), to 
guarantee that the potential drop at the edge is negligible even
without guard rings. Using the U.S. Naval Research Laboratory
facilities for laser scribing and ALD, diodes without guard rings and
a distance of only 14 $\mu$m between the active area and the cutting
edge were produced (Fig.~\ref{fig:active_edges}, right). 
These diodes show no evidence of breakdown up to 1 kV and a charge
collection efficiency close to 100\% even near the edge. 
In the future several tests are planned, in particular to assess the
radiation hardness of these technologies.

\section{Other activities}
The phase II ATLAS Inner Detector may exploit an additional pixel layer at larger radius in order
to improve tracking performances; some of the strip layers may also be replaced by pixels
to reduce the occupancy. The planar technology could be exploited since the expected fluences at 
those radii will not exceed a few 10$^{15}$ n$_{\rm eq}$/cm$^2$.
Because the instrumented area will be much larger than in the current pixel detector, cost efficiency 
becomes a major concern. On the sensor side, possible cost-saving measures 
include using 6" instead of 4" wafers and n-in-p instead of n-in-n sensors.
However, most of the cost of the ATLAS pixel detector was due to the bump bonding of 
the front-end electronic chips to the sensors. For this reason, an alternative 
technique based on Solid-Liquid InterDiffusion (SLID~\cite{bib:SLID}) 
has started to being investigated recently,
in collaboration with Fraunhofer Institute (IZM Munich). 
A (pixelated) thin (5~$\mu$m) layer of copper (Cu) is deposited through
mask electroplating on  both the sensors and the chip.
In addition, a 3~$\mu$m layer of tin (Sn) is overlaid on top of the Cu
on the sensor. 
The sensors and the chip surfaces are brought into contact in a
controlled environment where a temperature around 300 C and a pressure
around 5 bar is reached. At such temperatures the tin melts and an
eutectic alloy (Cu$_x$Sn$_y$) contact is formed. The alloy has a
melting temperature greater than 600 C and therefore does not liquify
during the process.
This technique should be cheaper than standard bump-bonding as it
requires less processing steps. It should also be exploitable with
small pitches and stacking. 
First results with thin (75 $\mu$m) n-in-p pixels connected to FEI3
chips showed promising results, with threshold dispersion, noise and
charge collection efficiency comparable to those of bump-bonded
sensors. Sporadic problems with chip misalignment and a few  
disconnected channels are under investigation.
\begin{figure}[!htbp]
\centering
\includegraphics[width=0.45\textwidth]{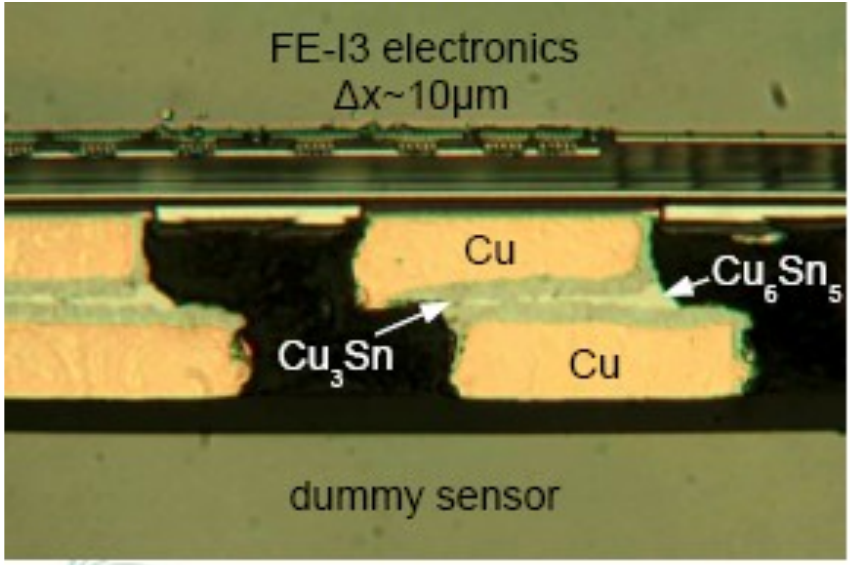}
\hspace{0.05\textwidth}
\includegraphics[width=0.45\textwidth]{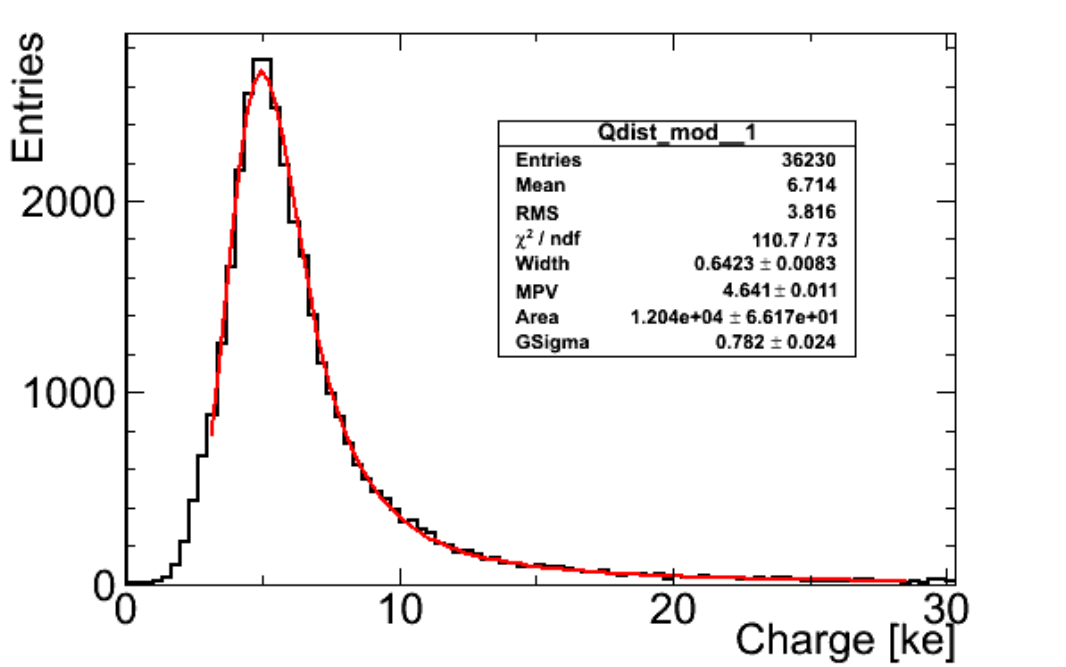}
\caption{Left: SLID bonding tests performed using a dummy sensor and a FEI3 chip. Right: 
charge collected from a MIP source with a 75~$\mu$m thick n-in-p sensor connected to a FEI3 chip using SLID.}
\label{fig:slid}
\end{figure}

\section{Conclusion}
Extensive R\&D by the PPS group has demonstrated that the proven planar technology of the 
ATLAS pixel detector is well suited even at the fluence ($5\times 10^{15}$ n$_{\rm eq}$/cm$^2$) 
predicted for the innermost layer in the phase I upgrade. The planar pixel sensors have been shown 
to operate and collect some charge even at phase II fluences ($2\times 10^{16}$ 
n$_{\rm eq}$/cm$^2$). Encouraging results have been obtained also
in terms of reduction of costs and of the inactive regions at the edge. Further activities are planned
in the future in order to evaluate the performances of thinner detectors and to establish the radiation 
hardness up to $2\times 10^{16}$ n$_{\rm eq}$/cm$^2$ of the various technologies under consideration.

\end{document}